\documentstyle[12pt]{article}

\let\ni=\noindent

\begin{document}

\baselineskip 0.8cm
 
\pagestyle {plain}

\setcounter{page}{1}

\renewcommand{\thefootnote}{\fnsymbol{footnote}}

\newcommand{\CKM}{Cabibbo---Kobayashi---Maskawa }

\newcommand{\UK}{Super--Kamiokande }

\newcommand{\onu }{\stackrel{\circ}{\nu} }

\newcommand{\mnu }{\nu_s^{(\mu)} }

\newcommand{\enu }{\nu_s^{(e)} }

\pagestyle {plain}

\setcounter{page}{1}

\pagestyle{empty}

~~~
\hfill IFT/99-08

\vspace{0.3cm}

\renewcommand{\thefootnote}{\fnsymbol{footnote}}

{\large\centerline{\bf Oscillations of the mixed pseudo--Dirac neutrinos%
\footnote{Supported in part by the Polish KBN--Grant 2 P03B 052 16 (1999--2000)
.}}}

\vspace{0.8cm}

{\centerline {\sc Wojciech Kr\'{o}likowski}}

\vspace{0.8cm}

{\centerline {\it Institute of Theoretical Physics, Warsaw University}}

{\centerline {\it Ho\.{z}a 69,~~PL--00--681 Warszawa, ~Poland}}

\vspace{0.5cm}

{\centerline{\bf Abstract}}

 Oscillations of three pseudo--Dirac flavor neutrinos $\nu_e $, $\nu_\mu $, $
\nu_\tau $ are considered: $0 < m^{(L)} = m^{(R)} \ll m^{(D)} $ for their 
Majorana and Dirac masses taken as universal before family mixing. The actual 
neutrino mass matrix is assumed to be the tensor product $ M^{(\nu)} \otimes {
\left( \begin{array}{cc} \lambda^{(L)} & 1 \\ 1 & \lambda^{(R)} \end{array} 
\right)}$, where $ M^{(\nu)}$ is a neutrino family mass matrix
($ M^{(\nu)\,\dagger} = M^{(\nu)}$) and $\lambda^{(L,R)} = m^{(L,R)}/m^{(D)}$. 
The $ M^{(\nu)}$ is tried in a form proposed previously for charged leptons $ 
e $, $\mu $, $\tau $ for which it gives $ m_\tau = 1776.80$ MeV {\it versus} $ 
m^{\rm exp}_\tau = 1777.05^{+0.29}_{-0.20}$ MeV (with the experimental values 
of $ m_e $ and $ m_\mu $ used as inputs). However, in contrast to the charged%
--lepton case, in the neutrino case its off--diagonal entries dominate over 
diagonal. Then, it is shown that three neutrino effects (the deficits of 
solar $\nu_e $'s and atmospheric $\nu_\mu $'s  as well as the possible LSND 
excess of $\nu_e $'s in accelerator $\nu_\mu $ beam) can be explained by 
neutrino oscillations though, alternatively, the LSND effect may be eliminated 
(by a parameter choice). Atmospheric $\nu_\mu $'s oscillate dominantly into 
$\nu_\tau $'s, while solar $\nu_e $'s --- into (automatically existing) 
Majorana sterile counterparts of $\nu_e $'s.

\vspace{0.2cm}

\ni PACS numbers: 12.15.Ff , 14.60.Pq , 12.15.Hh .

\vspace{0.5cm}

\ni April 1999

\vfill\eject

~~~~
\pagestyle {plain}

\setcounter{page}{1}

\vspace{0.1cm}

 Let us consider three flavor neutrinos $\nu_e $, $\nu_\mu $, $\nu_\tau $ and 
assume for them the mass matrix in the form of tensor product of the neutrino
family $ 3\times 3 $ mass matrix $ \left(M^{(\nu)}_{\alpha \beta}\right)\;\;
(\alpha,\beta = e,\,\mu,\,\tau)$ and the Majorana $2\times 2$ mass matrix

\vspace{-0.1cm}

\begin{equation}
\left( \begin{array}{cc} m^{(L)} & m^{(D)} \\ m^{(D)} & m^{(R)} \end{array} 
\right)\;\;,
\end{equation}

\ni the second divided by $ m^{(D)}$ (with $ m^{(D)}$ included into $ M^{(\nu)
}_{\alpha \beta}$). Then, the neutrino mass term in the lagrangian gets the 
form

\vspace{-0.1cm}

\begin{eqnarray}
-{\cal L}_{\rm mass} & = & \frac{1}{2}\sum_{\alpha \beta} \left(\overline{\onu
}^{(a)}_\alpha\;,\;\overline{\onu}^{(s)}_\alpha \right)\;{M}^{(\nu)}_{\alpha 
\beta} \; \left( \begin{array}{cc} \lambda^{(L)} & 1 \\ 1 & \lambda^{(R)} 
\end{array} \right)\;\left(\begin{array}{c} \onu^{(a)}_\beta \\ \onu_\beta^{(s)
} \end{array} \right) \nonumber \\ & = & \frac{1}{2}\sum_{\alpha \beta} \left(
\overline{\left( \onu_{\alpha L} \right)^c} \;,\; \overline{\onu_{\alpha R}} 
\right) \; {M}^{(\nu)}_{\alpha \beta}\;\left(\begin{array}{cc}\lambda^{(L)} & 1
\\ 1 & \lambda^{(R)} \end{array} \right)\;\left( \begin{array}{c} \onu_{\beta L
} \\ \left( \onu_{\beta R}\right)^c \end{array} \right) + {\rm h.c.}\;,
\end{eqnarray}

\vspace{-0.1cm}

\ni where

\vspace{-0.2cm}

\begin{equation}
\onu^{(a)}_\alpha \equiv\; \onu_{\alpha\,L} + \left(\onu_{\alpha\,L}\right)^c
\,,\,\onu^{(s)}_\alpha \equiv\; \onu_{\alpha\,R} + \left(\onu_{\alpha\,R}
\right)^c
\end{equation}

\ni and $\lambda^{(L,R)} \equiv m^{(L,R)}/m^{(D)}$. Here, $\onu^{(a)}_\alpha $ 
and $\onu_\alpha^{(s)}$ are the conventional Majorana active and sterile neutr%
inos of three families as they appear in the lagrangian before diagonalization 
of neutrino and charged--lepton family mass matrices. Due to the relation $
\overline{\nu^c_\alpha}\nu_\beta = \overline{\nu^c_\beta} \nu_\alpha $, the 
family mass matrix $ M^{(\nu)} = M^{(\nu)\,\dagger}$, when standing at the 
position of $\lambda^{(L)}$ and $\lambda^{(R)}$ in Eq. (2), reduces to its 
symmetric part $\frac{1}{2}(M^{(\nu)} + M^{(\nu)\,T} ) $ equal to its
real part $\frac{1}{2}(M^{(\nu)} + M^{(\nu)\,*}) = {\rm Re}\, M^{(\nu)} $. We 
will simply assume that (at least approximately) $ M^{(\nu)} = M^{(\nu)\,T} = 
M^{(\nu)\,*} $, and hence $ U^{(\nu)} = U^{(\nu)\,*} = \left(U^{(\nu)\,-1}
\right)^T $. Then, CP violation for neutrinos does not appear if, in addition, 
$ U^{(e)} = U^{(e)\,*} $. Further on, we will always assume that $ 0 <\lambda^{
(L)} = \lambda^{(R)}\;\; (\equiv \lambda^{(M)})$ and $\lambda^{(M)} \ll 1 $ 
(the pseudo--Dirac case) [1].

 Then, diagonalizing the neutrino mass matrix, we obtain from Eq. (2) 

\vspace{-0.1cm}

\begin{equation}
-{\cal L}_{\rm mass} = \frac{1}{2} \sum_i \left( \overline{\nu}^I_i\;,
\; \overline{\nu}_i^{II} \right)\;m_{\nu_i}\;\left( \begin{array}{cc} 
\lambda^I & 0 \\ 0 & \lambda^{II} \end{array} \right)\;\left(\begin{array}{c} 
\nu^{I}_i \\ \nu^{II}_i \end{array} \right)\;\;,
\end{equation}

\vspace{-0.1cm}

\ni where

\vspace{-0.2cm}

\begin{equation}
\left( {U}^{(\nu)\,\dagger}\right)_{i\,\alpha} {M}^{(\nu)}_{\alpha \beta} {U}^{
(\nu)}_{\beta\,j} = m_{\nu_i} \delta_{ij}\;\;\;,\;\;\;\lambda^{I,\,II} = \mp 1 
+ \lambda^{(M)} \simeq \mp 1
\end{equation}

\vspace{-0.2cm}

\ni $(i,\,j = 1,2,3)$ and

\vspace{-0.2cm}

\begin{equation}
\nu_i^{I,\,II} = \sum_i \left({U}^{(\nu)\,\dagger}\right)_{i\,\alpha} \frac{1}{
\sqrt{2}}\left(\onu^{(a)}_\alpha \mp \onu^{(s)}_\alpha\right) =
\sum_i {V}_{i\,\alpha} \frac{1}{\sqrt{2}}\left(\nu^{(a)}_\alpha \mp \nu^{(s)
}_\alpha\right)
\end{equation}

\vspace{-0.2cm}

\ni with $ V_{i\,\alpha} = \left({U}^{(\nu)\,\dagger}\right)_{i\,\beta} U^{(e)
}_{\beta \alpha}$ describing the lepton counterpart of the \CKM matrix. Here,

\vspace{-0.1cm}

\begin{equation}
\nu_\alpha^{(a,s)} \equiv \sum_\beta \left( U^{(e)\,\dagger}\right)_{\alpha 
\beta} \onu^{(a,s)}_\beta = \sum_i \left( V^\dagger \right)_{\alpha\,i} 
\frac{1}{\sqrt{2}}\left(\pm \nu^I_i + \nu^{II}_i \right) = \nu_{\alpha\,L,R} + 
\left(\nu_{\alpha\,L,R} \right)^c
\end{equation}

\vspace{-0.1cm}

\ni and

\vspace{-0.2cm}

\begin{equation}
\left( U^{(e)\,\dagger}\right)_{\alpha \gamma} M^{(e)}_{\gamma \delta} U^{(e)
}_{\delta \beta} = m_{e_\alpha} \delta_{\alpha \beta}\;,
\end{equation}

\ni where $ \left(M^{(e)}_{\alpha \beta}\right)\;\;(\alpha,\beta = e,\,\mu,\,
\tau)$ is the mass matrix for three charged leptons $e^-,\,\mu^-,\,\tau^- $, 
giving their masses $m_e,\,m_\mu,\,m_\tau $ after its diagonalization is 
carried out. Now, $\nu_\alpha^{(a)}$ and $\nu_\alpha^{(s)}$ are the convent%
ional Majorana active and sterile flavor neutrinos of three families, while $
\nu_i^I $ and $\nu_i^{II}$ are Majorana massive neutrinos.

 If CP violation for neutrinos does not appear or can be neglected, the 
probabilities for oscillations $\nu_\alpha^{(a)} \rightarrow \nu_\beta^{(a)}$ 
and $\nu_\alpha^{(a)} \rightarrow \nu_\beta^{(s)}$ are given by the following 
formulae (in the pseudo--Dirac case):

\vspace{-0.1cm}

\begin{eqnarray}
\lefteqn{P\left(\nu^{(a)}_\alpha \rightarrow \nu^{(a)}_\beta\right)\! =
\!|\langle \nu^{(a)}_\beta |e^{i P L}|\nu^{(a)}_\alpha \rangle |^2 = \delta_{
\beta\,\alpha} - \sum_i |V_{i\,\beta}|^2 |V_{i\,\alpha}|^2 \sin^2 \left( x^{II
}_i \! -\! x^I_i \right)} \nonumber \\ & & \!\!\! -\! \sum_{j>i} V_{j\,\beta} 
V^*_{j\,\alpha} V^*_{i\,\beta} V_{i\,\alpha}\!\left[ \sin^2 \left(x^I_j \! -\! 
x^I_i \right) + \sin^2 \left(x^{II}_j \! -\! x^{II}_i \right) + \sin^2 \left(
x^{II}_j\! - \! x^I_i \right) + \sin^2 \left( x^I_j \! -\! x^{II}_i \right) 
\right] \nonumber \\ & &
\end{eqnarray}

\vspace{-0.2cm}

\ni and

\vspace{-0.3cm}

\begin{eqnarray}
\lefteqn{P\left(\nu^{(a)}_\alpha \rightarrow \nu^{(s)}_\beta\right)\! =\!
|\langle \nu^{(s)}_\beta |e^{i P L}|\nu^{(a)}_\alpha \rangle |^2\! =\! \sum_i
|V_{i\,\beta}|^2 |V_{i\,\alpha}|^2 \sin^2 \left( x^{II}_i \! -\! x^I_i \right)}
\nonumber \\ & & \!\!\! -\!\sum_{j>i}\! V_{j\,\beta}V^*_{j\,\alpha}V^*_{i\,
\beta}V_{i\,\alpha}\left[ \sin^2 \left( x^I_j \! -\! x^I_i \right)\! + \!
\sin^2 \left(x^{II}_j \! -\! x^{II}_i \right)\! - \!\sin^2 \left( x^{II}_j\! 
-\! x^I_i \right)\! - \!\sin^2 \left(x^I_j\! -\! x^{II}_i \right) \right], 
\nonumber \\ & &
\end{eqnarray}

\ni where $ P |\nu^{I,\,II}_i \rangle = p^{I,\,II}_i |\nu^{I,\,II}_i \rangle 
$ , $p_i^{I,\,II} = \sqrt{E^2 - (m_{\nu_i}\lambda^{I,\,II})^2} \simeq E - 
(m_{\nu_i}\lambda^{I,\,II})^2/2E $ and

\vspace{-0.1cm}

\begin{equation}
x_i^{I,\,II} = 1.27\frac{ (m_{\nu_i}^2\lambda^{I,\,II})^2 L}{E}\;\;,\;\;
(\lambda^{I,\,II})^2 = 1 \mp 2\lambda^{(M)} \simeq 1
\end{equation}

\ni with $ m_{\nu_i}$, $ L $ and $ E $ expressed in eV, km and GeV, respect%
ively ($ L $ is the experimental baseline). Here, due to Eqs. (11),

\vspace{-0.1cm}

\begin{equation}
x_i^{II} - x_i^I = 1.27\frac{4 m_{\nu_i}^2\lambda^{(M)} L}{E}
\end{equation}

\ni and for $ j > i $

\vspace{-0.1cm}

\begin{equation}
x^I_j  - x^I_i  \simeq x^{II}_j  - x^{II}_i \simeq  x^{II}_j - x^I_i \simeq 
x^I_j - x^{II}_i \simeq 1.27\frac{( m_{\nu_j}^2 -  m_{\nu_i}^2) L}{E}\;.
\end{equation}

\ni Then, the bracket [ ] in Eq. (9) and (10) is reduced to $ 4 \sin^2 1.27
(m_{\nu_j}^2 - m_{\nu_i}^2) L/E $ and 0, respectively. The probability sum rule
$\sum_\beta \left[P\left(\nu^{(a)}_\alpha \rightarrow \nu^{(a)}_\beta\right) + 
P\left(\nu^{(a)}_\alpha \rightarrow \nu^{(s)}_\beta \right) \right] = 1 $ 
follows readily from Eqs. (9) and (10).

 Notice that in the case of lepton \CKM matrix being nearly unit, $\left( V_{i
\,\alpha}\right) \simeq \left(\delta_{i\,\alpha}\right)$, the oscillations $
\nu_\alpha^{(a)} \rightarrow \nu_\beta^{(a)}$ and $\nu_\alpha^{(a)} \rightarrow
\nu_\beta^{(s)}$ are essentially described by the formulae

\vspace{-0.1cm}

\begin{eqnarray}
P\left(\nu^{(a)}_\alpha \rightarrow \nu^{(a)}_\beta\right) & \simeq &
\delta_{\beta \alpha} - P\left(\nu^{(a)}_\alpha \rightarrow 
\nu^{(s)}_\beta \right) \;, \nonumber \\ P\left(\nu^{(a)}_\alpha \rightarrow 
\nu^{(s)}_\beta\right) & \simeq & \delta_{\beta \alpha} \sin^2 \left(
1.27\frac{4 m_{\nu_\alpha}^2\lambda^{(M)} L}{E} \right)  
\end{eqnarray}

\ni corresponding to three maximal mixings of $\nu^{(a)}_\alpha $ with $\nu^{(s
)}_\alpha $ $(\alpha = e,\,\mu,\,\tau)$. Of course, for a further discussion of
the oscillation formulae (9) and (10), in particular those for appearance modes
$\nu^{(a)}_\alpha \rightarrow \nu^{(a)}_\beta\;\;(\alpha \neq \beta)$, a 
detailed knowledge of $\left( V_{i\,\alpha}\right)$ is necessary.

 To this end, we will try to extend to neutrinos the form of charged--lepton 
mass matrix 

\vspace{-0.1cm}

\begin{equation}
\left({M}^{(e)}_{\alpha \beta}\right) = \frac{1}{29} \left(\begin{array}{ccc}
\mu^{(e)}\varepsilon^{(e)} & 2\alpha^{(e)} e^{i\varphi^{(e)}} & 0 \\ & & 
\\ 2\alpha^{(e)} e^{-i\varphi^{(e)}} & 4\mu^{(e)}(80 + \varepsilon^{(e)})/9 
& 8\sqrt{3}\,\alpha^{(e)} e^{i\varphi^{(e)}} \\ & & 
\\ 0 & 8\sqrt{3}\,\alpha^{(e)} e^{-i\varphi^{(e)}} & 24\mu^{(e)}
(624 + \varepsilon^{(e)})/25 \end{array}\right)
\end{equation}

\ni which reproduces surprisingly well the charged--lepton masses $m_e,\,m_\mu,
\,m_\tau $ ($\mu^{(e)}$, $\alpha^{(e)}$ and $\varepsilon^{(e)}$ are positive 
parameters). In fact, treating off--diagonal elements of $\left({M}^{(e)}_{
\alpha \beta}\right)$ as a perturbation of its diagonal entries, we get the 
mass sum rule

\begin{eqnarray}
m_\tau & = & \frac{6}{125}\left( 351 m_\mu - 136 m_e \right)
 + 10.2112 \left(\frac{\alpha^{(e)}}{\mu^{(e)}}
\right)^2 \;{\rm MeV} \nonumber \\
& = & \left[ 1776.80 + 10.2112 \left(\frac{\alpha^{(e)}}{\mu^{(e)}}
\right)^2\,\right]\;{\rm MeV}\;,
\end{eqnarray}

\ni where the experimental values of $ m_e $ and $ m_\mu $ are used as inputs. 
Then, $\mu^{(e)} = 85.9924$ MeV and $\varepsilon^{(e)} = 0.172329$ (up to the
perturbation). The prediction (16) agrees very well with the experimental 
figure $ m^{\rm exp}_\tau = 1777.05^{+0.29}_{-0.20}$ MeV, even in the zero 
order in $\left( \alpha^{(e)}/\mu^{(e)}\right)^2$. Taking this experimental 
value of $ m_\tau $ as another input, we obtain

\begin{equation}
\left(\frac{\alpha^{(e)}}{\mu^{(e)}}\right)^2 = 0.024^{+0.028}_{-0.025} \;,
\end{equation}

\ni what is not inconsistent with zero.

 Now, we conjecture the neutrino family mass matrix $\left( M^{(\nu)}_{\alpha 
\beta}\right)$ in the form (15) with $\mu^{(e)} \rightarrow \mu^{(\nu)}$, $
\alpha^{(e)} \rightarrow \alpha^{(\nu)}$, $\varepsilon^{(e)} \rightarrow 
\varepsilon^{(\nu)} \simeq 0 $ and $\varphi^{(e)} \rightarrow \varphi^{(\nu)} 
= 0 $ [2]. In order to get the neutrino family diagonalizing matrix $\left( U^{
(\nu)}_{\alpha\, i}\right)$ rather different from the unit matrix $\left(
\delta^{(\nu)}_{\alpha\,i}\right)$, we assume that diagonal elements of $\left(
M^{(\nu)}_{\alpha \beta}\right)$ can be considered as a perturbation of its 
off--diagonal entries (though the diagonal as well as the off--diagonal 
elements are expected to be very small). Under this assumption we derive the 
unitary matrix $\left( U^{(\nu)}_{\alpha\,i}\right)$ of the following form :

\begin{equation}
\left( U^{(\nu)}_{\alpha\,i}\right) = \left(\begin{array}{ccc} 
\frac{\sqrt{48}}{7} & -\frac{1}{7\sqrt{2}}e^{i\varphi^{(\nu)}} & \frac{1}{7
\sqrt{2}} e^{2i \varphi^{(\nu)}} \\ 0 &  \frac{1}{\sqrt{2}} & \frac{1}{\sqrt{
2}}\, e^{i\varphi^{(\nu)}} \\ -\frac{1}{7}\,e^{-2i \varphi^{(\nu)}} & 
-\frac{\sqrt{48}}{7\sqrt{2}}\, e^{-i\varphi^{(\nu)}} & \frac{\sqrt{48}}{7\sqrt{
2}} \end{array} \right) + O(\xi/7)
\end{equation}

\vspace{0.1cm}

\ni with $\varphi^{(\nu)} = 0 $ and

\begin{equation}
\xi \equiv \frac{M^{(\nu)}_{33}}{|M^{(\nu)}_{12}|} = 299.52\frac{\mu^{(\nu)
}}{\alpha^{(\nu)}}\;\;,\;\;\chi \equiv \frac{M^{(\nu)}_{22}}{|M^{(\nu)}_{
12}|} = \frac{\xi}{16.848} \;.
\end{equation}

\vspace{0.1cm}

\ni In this case, the neutrino family masses are

\vspace{-0.1cm}

\begin{equation}
m_{\nu_1} = \frac{\xi}{49}|M^{(\nu)}_{12}|\;\;,\;\; 
m_{\nu_2,\,\nu_3} = \left[ \mp 7 + \frac{1}{2}\left( \frac{48}{49} \xi + \chi
\right)\right] |M^{(\nu)}_{12}| \;,
\end{equation}

\ni where $|M^{(\nu)}_{12}| = 2\alpha^{(\nu)}/29 $ (thus, $m_{\nu_1} \ll |m_{
\nu_2}| < m_{\nu_3}$). Hence,

\vspace{-0.1cm}

\begin{equation}
m_{\nu_3}^2 - m_{\nu_2}^2 = 14\left(\frac{48}{49}\xi + \chi\right)|M^{(\nu)}_{
12}|^2 = 20.721\, \alpha^{(\nu)} \mu^{(\nu)} \;.
\end{equation}

\ni Taking in contrast $\left( U^{(e)}_{\alpha \beta}\right) \simeq \left(
\delta_{\alpha \beta}\right)$ --- as in $\left(M^{(e)}_{\alpha \beta}
\right)$ the off--diagonal elements are perturbatively small {\it versus} 
diagonal entries [{\it cf.} Eq. (17)] --- we can insert

\vspace{-0.1cm}

\begin{equation}
V_{i\,\alpha} \simeq \left( U^{(\nu)\,\dagger}\right)_{i\,\alpha} = U^{(\nu)\,*
}_{\alpha\,i}
\end{equation}

\ni into Eqs. (9) and (10). Here, $U^{(\nu)\,*}_{\alpha\,i}$ are determined 
from Eq. (18). 

 Then, with the use of Eqs. (12) and (13) the $\nu^{(a)}_\alpha \rightarrow 
\nu^{(a)}_\beta $ oscillation formulae (9) take the form

\vspace{-0.1cm}

\begin{eqnarray}
P\left(\nu^{(a)}_e \rightarrow \nu^{(a)}_e \right) & = & 1 - \frac{48^2}{49^2}
\sin^2 \left( 1.27 \frac{4 m_{\nu_1}^2\lambda^{(M)} L}{E} \right) \nonumber \\ 
& &  - \frac{1}{4\cdot 49^2}\left[ \sin^2 \left( 1.27 \frac{4 m_{\nu_2}^2
\lambda^{(M)} L}{E} \right) + \sin^2 \left( 1.27 \frac{4 m_{\nu_3}^2\lambda^{(
M)} L}{E} \right) \right] \nonumber \\ & & 
- \frac{96}{49^2}\left[\sin^2 \left( 1.27 \frac{(m_{\nu_2}^2 - m^2_{\nu_1})
L}{E} \right) + \sin^2 \left( 1.27 \frac{(m_{\nu_3}^2 - m_{\nu_1}^2) L}{E} 
\right) \right] \nonumber \\ & & - \frac{1}{49^2} \sin^2 \left( 1.27 
\frac{(m_{\nu_3}^2 - m_{\nu_2}^2) L}{E} \right)\;, \nonumber \\ 
P\left( \nu^{(a)}_\mu \rightarrow \nu^{(a)}_\mu \right) & = & 1 - \frac{1}{4}
\left[\sin^2 \left( 1.27 \frac{4 m_{\nu_2}^2\lambda^{(M)} L}{E} \right) + 
\sin^2 \left( 1.27 \frac{4 m_{\nu_3}^2\lambda^{(M)} L}{E} \right)\right]
\nonumber \\ & & - \sin^2 \left( 1.27 \frac{(m_{\nu_3}^2 - m^2_{\nu_2}) L}{E} 
\right) \;, \nonumber \\ 
P\left( \nu^{(a)}_\mu \rightarrow \nu^{(a)}_e \right) & = &  - \frac{1}{4\cdot
49} \left[ \sin^2 \left( 1.27 \frac{4 m_{\nu_2}^2\lambda^{(M)} L}{E} \right) + 
\sin^2 \left( 1.27 \frac{4 m_{\nu_3}^2\lambda^{(M)} L}{E} \right)\right]
\nonumber \\ & & + \frac{1}{49} \sin^2 \left( 1.27\frac{(m_{\nu_3}^2 - m^2_{
\nu_2}) L}{E} \right) 
\end{eqnarray}

\ni and

\vspace{-0.2cm}

\begin{eqnarray}
P\left(\nu^{(a)}_\mu \rightarrow \nu^{(a)}_\tau \right) & = & - \frac{48}{4
\cdot 49} \left[\sin^2 \left( 1.27 \frac{4 m_{\nu_2}^2\lambda^{(M)} L}{E} 
\right) + \sin^2 \left( 1.27 \frac{4 m_{\nu_3}^2\lambda^{(M)} L}{E} \right)
\right] \nonumber \\ & & + \frac{48}{49} \sin^2 \left( 1.27 \frac{(m_{\nu_3}^2 
- m^2_{\nu_2}) L}{E} \right) \;, \nonumber \\ 
P\left(\nu^{(a)}_e \rightarrow \nu^{(a)}_\tau \right) & = & - \frac{48}{49^2}
\sin^2 \left( 1.27 \frac{4 m_{\nu_1}^2\lambda^{(M)} L}{E} \right) \nonumber \\ 
& & - \frac{48}{4\cdot 49^2}\left[ \sin^2 \left( 1.27 \frac{4 m_{\nu_2}^2
\lambda^{(M)} L}{E} \right) + \sin^2 \left( 1.27 \frac{4 m_{\nu_3}^2\lambda^{(
M)} L}{E} \right) \right] \nonumber \\   
& & + \frac{96}{49^2}\left[ \sin^2 \left( 1.27 \frac{(m_{\nu_2}^2 - m^2_{\nu_1
}) L}{E} \right) + \sin^2 \left( 1.27 \frac{(m_{\nu_3}^2 - m_{\nu_1}^2) L}{E} 
\right) \right] \nonumber \\ & & - \frac{48}{49^2} \sin^2 \left( 1.27 
\frac{(m_{\nu_3}^2 - m_{\nu_2}^2) L}{E} \right)\;, \nonumber \\ 
P\left(\nu^{(a)}_\tau \rightarrow \nu^{(a)}_\tau \right) & = & 1 - \frac{1}{
49^2} \sin^2 \left( 1.27 \frac{4 m_{\nu_1}^2\lambda^{(M)} L}{E} \right) 
\nonumber \\ & & - \frac{48^2}{4\cdot 49^2}\left[ \sin^2 \left( 1.27 \frac{4 
m_{\nu_2}^2 \lambda^{(M)} L}{E} \right) + \sin^2 \left( 1.27 \frac{4 m_{
\nu_3}^2\lambda^{(M)} L}{E} \right) \right] \nonumber \\   
& & - \frac{96}{49^2}\left[\sin^2 \left( 1.27 \frac{(m_{\nu_2}^2 - m^2_{\nu_1})
L}{E} \right) + \sin^2 \left( 1.27 \frac{(m_{\nu_3}^2 - m_{\nu_1}^2) L}{E} 
\right) \right] \nonumber \\ & & - \frac{48^2}{49^2} \sin^2 \left( 1.27 
\frac{(m_{\nu_3}^2 - m_{\nu_2}^2) L}{E} \right)\;.
\end{eqnarray}

\ni In these formulae, the experimental baselines $ L $ (and neutrino energies
$ E $) are generally different.

 Further on, we intend to relate the first, second and third Eq. (23) to 
the experimental results concerning the deficit of solar $\nu_e $'s [3], the 
deficit of atmospheric $\nu_\mu $'s [4] and the excess of $\nu_e $'s in 
accelerator $\nu_\mu $ beam [5], respectively.

 First, let us assume the simplifying hypothesis that the LSND effect [5] does
not exist. Then, under the numerical conjecture that

\vspace{-0.1cm}

\begin{eqnarray}
1.27 \frac{4 m_{\nu_1}^2\lambda^{(M)} L_{\rm sol}}{E_{\rm sol}}\;\, = O(1)\;\;
\;,\;\;\;\;1.27 \frac{4 m_{\nu_2}^2\lambda^{(M)} L_{\rm atm}}{E_{\rm atm}} & = 
& O \left(\frac{m_{\nu_2}^2 L_{\rm atm}/E_{\rm atm}}{m_{\nu_1}^2 L_{\rm sol}
/E_{\rm sol}}\right)\ll 1\,, \nonumber \\ 1.27 \frac{(m_{\nu_3}^2 - m_{\nu_2}^2
) L_{\rm atm}}{E_{\rm atm}} = O(1)\;\;\;,\;\,
1.27 \frac{(m_{\nu_3}^2 - m_{\nu_2}^2) L_{\rm sol}}{E_{\rm sol}} & = & O\left(
\frac{L_{\rm sol}/E_{\rm sol}}{L_{\rm atm}/E_{\rm atm}}\right) \gg 1 \,, 
\end{eqnarray}

\ni we obtain from Eqs. (23)

\vspace{-0.1cm}

\begin{eqnarray}
P\left(\nu^{(a)}_e \rightarrow \nu^{(a)}_e \right) & \simeq & 1 - \frac{48^2}{
49^2} \sin^2 \left( 1.27 \frac{4 m_{\nu_1}^2\lambda^{(M)} L_{\rm sol}}{E_{\rm 
sol}} \right) - \frac{387}{4\cdot 49^2} \nonumber \\ & \simeq &  1 - \frac{
48^2}{49^2} \sin^2 \left( 1.27 \frac{4 m_{\nu_1}^2\lambda^{(M)} L_{\rm sol}}{
E_{\rm sol}} \right)\;, \nonumber \\ 
P\left( \nu^{(a)}_\mu \rightarrow \nu^{(a)}_\mu \right) & \simeq & 1 - \sin^2 
\left( 1.27 \frac{(m_{\nu_3}^2 - m_{\nu_2}^2) L_{\rm atm}}{E_{\rm atm}} \right)
- 8(1.27)^2 \frac{m_{\nu_2}^4\lambda^{(M)\,2} L_{\rm atm}^2}{E^2_{\rm 
atm}} \nonumber \\ & \simeq & 1 - \sin^2 \left( 1.27 \frac{(m_{\nu_3}^2 - 
m_{\nu_2}^2) L_{\rm atm}}{E_{\rm atm}} \right) \;\,, \nonumber \\
P\left( \nu^{(a)}_\mu \rightarrow \nu^{(a)}_e \right) & \simeq & -\frac{8}{49} 
(1.27)^2 \frac{m_{\nu_2}^4 \lambda^{(M)\,2} L_{\rm LSND}^2}{E_{\rm LSND}^2} 
+ \frac{1}{49}(1.27)^2 \frac{(m_{\nu_3}^2 - m_{\nu_2}^2)^2 L_{\rm LSND}^2}{
E_{\rm LSND}^2} \nonumber \\ & \simeq & 0 \;\,.
\end{eqnarray}

\ni The term $-387/4\cdot 49^2 = -0.0403 $ in the first Eq. (26) comes out 
from averaging all $\sin^2 $ of large phases over oscillation lengths defined 
by $\sin^2 $ of a phase $= O(1)$ (then, each $\sin^2 $ of a large phase gives
1/2). 

 Comparing Eqs. (26) with experimental estimates, we get for solar $\nu_e $'s 
[3] (using the global vacuum fit)

\vspace{-0.1cm}

\begin{equation}
\frac{48^2}{49^2} \leftrightarrow \sin^2 2\theta_{\rm sol} \sim 0.75 \;\;,\;\; 
4 m_{\nu_1}^2 \lambda^{(M)} \leftrightarrow \Delta m^2_{\rm sol} \sim 6.5
\times 10^{-11}\;{\rm eV}^2 \;,
\end{equation}

\ni and for atmospheric $\nu_\mu $'s [4]

\vspace{-0.1cm}

\begin{equation}
1 \leftrightarrow \sin^2 2\theta_{\rm atm} \sim 1\;\;,\;\;m_{\nu_3}^2 - m_{
\nu_2}^2 \leftrightarrow \Delta m^2_{\rm atm} \sim 2.2\times 10^{-3}\;{\rm 
eV}^2\;.
\end{equation}

\ni Thus, from Eqs. (27) and (28)

\vspace{-0.1cm}

\begin{equation}
\frac{4 m_{\nu_1}^2 \lambda^{(M)}}{m_{\nu_3}^2 - m_{\nu_2}^2} \leftrightarrow 
\frac{\Delta m_{\rm sol}^2}{\Delta m_{\rm atm}^2} \sim 3.0\times 10^{-8}\;.
\end{equation}

\ni Hence, making use of Eqs. (20) and (21), we infer that

\begin{eqnarray}
\xi \lambda^{(M)} \sim 2.6\times 10^{-4}\; & , & \;\; \frac{m_{\nu_1}}{
|m_{\nu_2}|} \lambda^{(M)} = \frac{1}{7^3} \xi \lambda^{(M)} \sim 7.5 \times
10^{-7} \;\;, \nonumber \\ \frac{\mu^{(\nu)}}{\alpha^{(\nu)}} \lambda^{(M)} = 
\frac{1}{299.52} \xi \lambda^{(M)} \sim 8.6 \times 10^{-7} \; & , & \;\;
\alpha^{(\nu)} \mu^{(\nu)} = \frac{m_{\nu_3}^2 - m_{\nu_2}^2}{20.721} \sim 1.1
\times 10^{-4}\;{\rm eV}^2 \;\;, \nonumber \\ \mu^{(\nu)\,2} \lambda^{(M)} \sim
9.1 \times 10^{-11}\;{\rm eV}^2 \; & , & \;\;\frac{\alpha^{(\nu)\,2}}{
\lambda^{(M)}} \sim 1.2 \times 10^2 \;{\rm eV}^2\;\;.
\end{eqnarray}

\ni Here, the constant $\xi $ still may be treated as a free parameter (deter%
mining $\lambda^{(M)}$). If $\xi = O(10^{-1})$, then $\lambda^{(M)} 
= O(10^{-3})$, $ m_{\nu_1}/|m_{\nu_2}| = O(10^{-4})$, $\mu^{(\nu)}/\alpha^{(
\nu)} = O(10^{-4})$, $\mu^{(\nu)} = O(10^{-4}\;{\rm eV})$, $\alpha^{(\nu)}  = 
O(1\;{\rm eV})$ and 

\begin{equation}
m_{\nu_1} = O(10^{-4}\;{\rm eV})\;\;,\;\; |m_{\nu_2}| = O(10^{-1}\;{\rm eV})
\;\;,\;\; m_{\nu_3} = O(10^{-1}\;{\rm eV})
\end{equation}

\ni with $m_{\nu_3}^2 - m_{\nu_2}^2 \sim 2.2 \times 10^{-3}\;{\rm eV}^2$.

 In this way, both neutrino deficits can be explained by pseudo--Dirac neutrino
oscillations. Note that solar $\nu^{(a)}_e$'s and atmospheric $\nu^{(a)}_\mu 
$'s oscillate dominantly into $\nu^{(s)}_e $'s and $\nu^{(a)}_\tau $'s, res%
pectively (here, $\nu^{(a)}_{\alpha\,L} = \nu_{\alpha\,L}$, $\nu^{(s)}_{\alpha
\,L} = (\nu_{\alpha}^c)_L $ ).

 Now, let us accept the LSND effect [5]. Then, making the numerical conjecture 
that

\begin{eqnarray}
1.27 \frac{4 m_{\nu_1}^2\lambda^{(M)} L_{\rm sol}}{E_{\rm sol}}\;\, = O(1)\;\;
\,,\,\;\;\;\;1.27 \frac{4 m_{\nu_2}^2\lambda^{(M)} L_{\rm atm}}{E_{\rm atm}} & 
= & O \left( \frac{m_{\nu_2}^2 L_{\rm atm}/E_{\rm atm}}{m_{\nu_1}^2 L_{\rm 
sol}/E_{\rm sol}}\right) <1\,, \nonumber \\ 1.27 \frac{(m_{\nu_3}^2 - m_{
\nu_2}^2) L_{\rm LSND}}{E_{\rm LSND}} = O(1)\;\;,\;
1.27 \frac{(m_{\nu_3}^2 - m_{\nu_2}^2) L_{\rm atm}}{E_{\rm atm}} & = & O\left(
\frac{L_{\rm atm}/E_{\rm atm}}{L_{\rm LSND}/E_{\rm LSND}}\right) \gg 1\,, 
\nonumber \\ & & 
\end{eqnarray}

\ni we get from Eqs. (23)

\begin{eqnarray}
P\left(\nu^{(a)}_e \rightarrow \nu^{(a)}_e \right) & \simeq & 1 - \frac{48^2}{
49^2} \sin^2 \left( 1.27 \frac{4 m_{\nu_1}^2\lambda^{(M)} L_{\rm sol}}{E_{\rm 
sol}} \right) - \frac{387}{4\cdot 49^2} \nonumber \\ & \simeq &  1 - \frac{
48^2}{49^2} \sin^2 \left( 1.27 \frac{4 m_{\nu_1}^2\lambda^{(M)} L_{\rm sol}}{
E_{\rm sol}} \right)\;, \nonumber \\ 
P\left( \nu^{(a)}_\mu \rightarrow \nu^{(a)}_\mu \right) & \simeq & 1 - \frac{1
}{2} - \frac{1}{2} \sin^2 \left( 1.27 \frac{4 m_{\nu_2}^2 \lambda^{(M)} 
L_{\rm atm}}{E_{\rm atm}} \right)\;, \nonumber \\ 
P\left( \nu^{(a)}_\mu \rightarrow \nu^{(a)}_e \right) & \simeq & \frac{1}{49} 
\sin^2 \left( 1.27 \frac{(m_{\nu_3}^2 - m_{\nu_2}^2) L_{\rm LSND}}{E_{\rm LSND}}
\right) - \frac{8}{49} (1.27)^2 \frac{m_{\nu_2}^4 \lambda^{(M)\,2} L_{\rm LSND
}^2}{E_{\rm LSND}^2} \nonumber \\ & \simeq & \frac{1}{49} \sin^2 \left( 1.27 
\frac{(m_{\nu_3}^2 - m_{\nu_2}^2) L_{\rm LSND}}{E_{\rm LSND}} \right) \;\,.
\end{eqnarray}

 When comparing Eqs. (33) with experimental estimates, we obtain for solar $
\nu_e $'s [3] (making use of global vacuum fit)

\begin{equation}
\frac{48^2}{49^2} \leftrightarrow \sin^2 2\theta_{\rm sol} \sim 0.75 \;\;,\;\; 
4 m_{\nu_1}^2 \lambda^{(M)} \leftrightarrow \Delta m^2_{\rm sol} \sim 6.5 
\times 10^{-11}\;{\rm eV}^2 \;,
\end{equation}

\ni for atmospheric $\nu_\mu $'s [4]

\vspace{-0.1cm}

\begin{equation}
1 \leftrightarrow \sin^2 2\theta_{\rm atm} \sim 1\;\;,\;\;\frac{1}{2} + 
\frac{1}{2} \sin^2 \left( 1.27 \frac{4 m_{\nu_2}^2 \lambda^{(M)} 
L_{\rm atm}}{E_{\rm atm}} \right) \leftrightarrow \sin^2 \left( 1.27 
\frac{\Delta m_{\rm atm}^2 L_{\rm atm}}{E_{\rm atm}} \right)
\end{equation}

\ni with

\begin{equation}
\Delta m^2_{\rm atm} \sim 2.2\times 10^{-3}\;{\rm eV}^2\;,
\end{equation}

\ni and for accelerator $\nu_\mu $'s [5], say,

\begin{equation}
\frac{1}{49} \leftrightarrow \sin^2 2\theta_{\rm LSND} \sim 0.02 \;\;,\;\; 
m_{\nu_3}^2 - m_{\nu_2}^2 \leftrightarrow \Delta m_{\rm LSND}^2 \sim 0.5 \;
{\rm eV}^2 \;.
\end{equation}

\ni So, from Eqs. (34) and (37)

\begin{equation}
\frac{4 m_{\nu_1}^2 \lambda^{(M)}}{m_{\nu_3}^2 - m_{\nu_2}^2} \leftrightarrow 
\frac{\Delta m_{\rm sol}^2}{\Delta m_{\rm LSND}^2} \sim 1.3\times 10^{-10}\;.
\end{equation}

\ni Hence, due to Eqs. (20) and (21),

\begin{eqnarray}
\xi \lambda^{(M)} \sim 1.1\times 10^{-6}\;\; & , & \;\; \frac{m_{\nu_1}}{
|m_{\nu_2}|} \lambda^{(M)} \sim 3.3 \times 10^{-9} \;\;, \nonumber \\ 
\frac{\mu^{(\nu)}}{\alpha^{(\nu)}} \lambda^{(M)} \sim 3.8 \times 10^{-9} \;\; 
& , & \;\; \alpha^{(\nu)} \mu^{(\nu)} \sim 2.4 \times 10^{-2}\;{\rm eV}^2 
\;\;, \nonumber \\ \mu^{(\nu)\,2} \lambda^{(M)} \sim 9.2 \times 10^{-11} \;
{\rm eV}^2 \;\; & , & \;\;\frac{\alpha^{(\nu)\,2}}{\lambda^{(M)}} \sim 6.3 
\times 10^6 \;{\rm eV}^2 \;\;.
\end{eqnarray}

\ni Here, the constant $\xi $ still may play the role of a free parameter 
(determining $\lambda^{(M)} $). If $\xi = O(10^{-1})$, then $\lambda^{(M)} = 
O(10^{-5})$, $ m_{\nu_1}/|m_{\nu_2}| = O(10^{-4})$, $\mu^{(\nu)}/\alpha^{(\nu)}
= O(10^{-4})$, $\mu^{(\nu)} = O(10^{-3}\;{\rm eV})$, $\alpha^{(\nu)} =
O(10\;{\rm eV})$, and hence

\begin{equation}
m_{\nu_1} = O(10^{-3}\;{\rm eV})\;\;,\;\;|m_{\nu_2}| = O(1\;{\rm eV})\;\;,
\;\;m_{\nu_3} = O(1\;{\rm eV})
\end{equation}

\ni with $m_{\nu_3}^2 - m_{\nu_2}^2 \sim 0.5\;{\rm eV}^2 $. Then, in Eq. (35) 
we can put approximately

\begin{equation}
\frac{1}{2} \leftrightarrow \sin^2 \left( 1.27 \frac{\Delta m_{\rm atm}^2 
L_{\rm atm}}{E_{\rm atm}} \right) \simeq 1 - U/D \sim 1 - 
0.54^{+0.06}_{-0.05}
\end{equation}

\ni in a reasonable consistency with the \UK estimate [4]. Here, $(U-D)/(U+D)$
is the up--down assymetry for $\nu_\mu $'s, estimated as $-0.296 \pm 0.048 \pm 
0.01 $.

 In this way, therefore, all three neutrino effects can be explained by pseudo%
--Dirac neutrino oscillations. Note that solar $\nu_e^{(a)}$'s and atmospheric
$\nu_\mu^{(a)}$'s oscillate dominantly into $\nu_e^{(s)}$'s and $\nu_\tau^{(a)
}$'s, respectively, as in the previous case when the LSND effect was absent.

 The recently improved upper bound on the effective mass $ \langle m_{\nu_e} 
\rangle $ of the Majorana $\nu_e^{(a)}$ neutrino extracted from neutrinoless 
double $\beta $ decay experiments is $ 0.2 $ eV [6]. In our pseudo--Dirac case,
this mass is given by the formula (if $V_{i \alpha} \simeq U^{(\nu)\,*}_{
\alpha i}$):

\begin{eqnarray}
\langle m_{\nu_e} \rangle & = & |\sum_i U^{(\nu)\,2}_{e\,i} m_{\nu_i} \frac{1
}{2} \left(\lambda^{I} + \lambda^{II}\right)| = |\sum_i U^{(\nu)\,2}_{e\,i} 
m_{\nu_i}\lambda^{(M)}| \nonumber \\ & = & \frac{1}{49\cdot 29}\left(3\cdot 
\frac{48}{49} \xi + \chi \right) \alpha^{(\nu)} \lambda^{(M)} = \frac{\xi 
\lambda^{(M)}}{473.96} \alpha^{(\nu)}\;,
\end{eqnarray}

\ni as $\varphi^{(\nu)} = 0 $ in Eq. (18) (here, $U_{\alpha i} = U^{*}_{\alpha 
i}$) and $\lambda^{I,\,II} = \mp 1 + \lambda^{(M)}$, while 

\begin{equation}
\nu^{(a)}_\alpha = \sum_i U^{(\nu)}_{\alpha i} \frac{1}{\sqrt{2}}\left(\nu_i^
{I} + \nu_i^{II}\right)\;.
\end{equation}

\ni Thus, in the option excluding or accepting LSND effect we estimate from Eq.
(30) or (39) that

\begin{equation}
\langle m_{\nu_e} \rangle \sim \left\{ \begin{array}{l} 5.4 \times 10^{-7}
\alpha^{(\nu)} \sim O(10^{-6}\;{\rm eV}) \\ 2.4 \times 10^{-9} \alpha^{(\nu)} 
\sim O(10^{-8}\;{\rm eV}) \end{array} \right.\;,
\end{equation}

\ni respectively. Thus, in this pseudo--Dirac case, the $0\nu\beta\beta $ decay
violating the lepton number conservation is negligible. Note that $\langle m_{
\nu_e} \rangle \ll m_{\nu_1} \ll |m_{\nu_2}| < m_{\nu_3}$ in both options. 
Here, the neutrino masses are

\begin{equation}
m_{\nu_i}^{I,\,II} = m_{\nu_i} \lambda^{I,\,II} = m_{\nu_i}\left(\mp 1 + 
\lambda^{(M)}\right) \simeq \mp m_{\nu_i}\;.
\end{equation}

\ni Since for relativistic particles only masses squared are relevant, the 
"phenomenological" neutrino masses are equal to $ |m_{\nu_i}^{I,\,II}| \simeq 
|m_{\nu_i}|$ {\it i.e.}, $\simeq m_{\nu_1}\;,\;|m_{\nu_2}|\;,\;m_{\nu_3}$.

 Finally, let us turn back to the option, where there is no LSND effect. In 
this case, the natural possibility seems to be a (nearly) diagonal form of 
neutrino family mass matrix $ M^{(\nu)} \simeq \left( \delta_{\alpha \beta}\,
m_{\nu_\alpha}\right)$ and so, unit neutrino diagonalizing matrix $ U^{(\nu)} 
\simeq \left( \delta_{\alpha i}\right)$. Then, if $ U^{(e)} \simeq \left( 
\delta_{\alpha \beta}\right)$ {\it i.e.}, $ V \simeq \left( \delta_{i \alpha}
\right)$, Eqs. (14) hold, giving

\begin{eqnarray}
P\left(\nu^{(a)}_e \rightarrow \nu^{(a)}_e \right) & \simeq & 1 - P\left(
\nu^{(a)}_e \rightarrow \nu^{(s)}_e \right) \simeq  1 - \sin^2 \left( 1.27 
\frac{4 m_{\nu_e}^2\lambda^{(M)} L}{E} \right)\;, \nonumber \\ 
P\left( \nu^{(a)}_\mu \rightarrow \nu^{(a)}_\mu \right) & \simeq & 1 - 
P\left( \nu^{(a)}_\mu \rightarrow \nu^{(s)}_\mu \right) \simeq 1 - \sin^2 \left(
1.27 \frac{4 m^2_{\nu_\mu} \lambda^{(M)} L}{E} \right)\;. 
\end{eqnarray}

\ni Here, $m_{\nu_i} = m_{\nu_\alpha}$ are neutrino family masses.

Comparing Eqs. (46) with experimental estimates for solar $\nu_e $'s [3] (using
the global vacuum fit) and atmospheric $\nu_\mu $'s [4], we have Eq. (27) (with
$m_{\nu_i} = m_{\nu_e}$) and the relation

\begin{equation}
1 \leftrightarrow \sin^2 2\theta_{\rm atm} \sim 1\;\;,\;\; 
4 m_{\nu_\mu}^2 \lambda^{(M)} \leftrightarrow \Delta m_{\rm atm}^2 
\sim 2.2 \times 10^{-3} \;{\rm eV}^2\;\;,
\end{equation}

\ni respectively. Hence,

\begin{equation}
\frac{m_{\nu_e}^2}{m_{\nu_\mu}^2} \leftrightarrow \frac{\Delta m_{\rm sol}^2}
{\Delta m_{\rm atm}^2} \sim 3.0\times 10^{-8}\;.
\end{equation}

\ni Under the conjecture that $M^{(\nu)}$ has the form (15) with $\mu^{(e)} 
\rightarrow \mu^{(\nu)}$, $\alpha^{(e)} \rightarrow \alpha^{(\nu)} = 0 $, 
$\varepsilon^{(e)} \rightarrow \varepsilon^{(\nu)} \simeq 0$, we get

\begin{equation}
\frac{m_{\nu_e}}{m_{\nu_\mu}} = \frac{9\, \varepsilon^{(\nu)}}{4\cdot 80} 
\leftrightarrow \left(\frac{\Delta m_{\rm sol}^2}{\Delta m_{\rm atm}^2}\right)^{
1/2} \sim 1.7 \times 10^{-4}\;.
\end{equation}

\ni Then,

\vspace{-0.2cm}

\begin{equation}
\varepsilon^{(\nu)} \sim 6.1\times 10^{-3}\;
\end{equation}

\vspace{-0.1cm}

\ni and 

\vspace{-0.2cm}

\begin{eqnarray}
m_{\nu_e} & = & \frac{\varepsilon^{(\nu)}}{29} \mu^{(\nu)} \sim 2.1\times 10^{
-4}\;\mu^{(\nu)}\;\,,\nonumber \\
m_{\nu_\mu} = \frac{4\cdot 80}{9\cdot 29}\,\mu^{(\nu)} = 1.2261\,\mu^{(\nu)} 
& , & m_{\nu_\tau} = \frac{24\cdot 624}{25\cdot 29}\,\mu^{(\nu)} 
= 20.657\;\mu^{(\nu)} = 16.848 m_{\nu_\mu}\;\,. \nonumber \\ & &
\end{eqnarray}

\vspace{-0.1cm}

\ni Here, the neutrino masses are $ m_{\nu_\alpha}^{I,\,II} = m_{\nu_\alpha}
\lambda^{I,\,II} = m_{\nu_\alpha}(\mp 1 + \lambda^{(M)}) \simeq \mp m_{
\nu_\alpha}$, so that $| m_{\nu_\alpha}^{I,\,II}| \simeq m_{\nu_e}\,,\;
m_{\nu_\mu}\,,\;m_{\nu_\tau}\,$, where $m_{\nu_e}\,:\;m_{\nu_\mu}\,:\;
m_{\nu_\tau} \sim 1.7 \times 10^{-4} : 1 : 16.8 $. From Eqs. (47) and (51) 
we infer that

\begin{equation}
\mu^{(\nu)\,2} \lambda^{(M)}  \sim 3.7 \times 10^{-4}\;{\rm eV}^2\;.
\end{equation}

 In this way, both neutrino deficits can be explained by oscillations of un%
mixed pseudo--Dirac neutrinos ($ U^{(\nu)}_{\alpha i} \simeq \delta_{\alpha i}
$). Note, however, that now both solar $\nu^{(a)}_e $'s and atmospheric $\nu^{(
a)}_\mu $'s oscillate dominantly into Majorana sterile neutrinos: $\nu^{(s)}_e 
$'s and $\nu^{(s)}_\mu $'s, respectively (in contrast to the previous mixed 
pseudo--Dirac $\nu^{(a)}_e $ and $\nu^{(a)}_\mu $ neutrinos of which the latter
oscillated dominantly into $\nu^{(a)}_\tau $'s). The experimental evidence for 
$\nu_\mu \rightarrow \nu_\tau $ oscillations and/or for the LSND effect would 
be, of course, crucial in the process of understanding the mechanism of 
neutrino oscillations.

 In the present case, the effective mass $\langle m_{\nu_e} \rangle $ of the 
Majorana $\nu^{(a)}_e $ neutrino is given as

\begin{equation}
\langle m_{\nu_e} \rangle \simeq m_{\nu_e} \frac{1}{2} \left(\lambda^{I} + 
\lambda^{II}\right) = m_{\nu_e}\lambda^{(M)} \;, 
\end{equation}

\ni since $ U^{(\nu)}_{\alpha i} \simeq \delta_{\alpha i}$. Thus, the $O \nu 
\beta \beta $ decay upper bound $\langle m_{\nu_e} \rangle \leq 0.2$ eV is 
certainly satisfied because of $\lambda^{(M)} \ll 1 $ (and $ m_{\nu_e} \leq $ a
few eV).

 If it turned out that both solar $\nu_e $'s and atmospheric $\nu_\mu $'s 
oscillated into sterile neutrinos, it would not be easy to recognize whether, 
as discussed above, the latter should be Majorana sterile counterparts of 
Majorana active $\nu_e $'s and $\nu_\mu $'s, or rather, two extra Dirac sterile
neutrinos [7].

\vfill\eject 

~~~~
\vspace{0.6cm}

{\bf References}

\vspace{1.0cm}

{\everypar={\hangindent=0.5truecm}
\parindent=0pt\frenchspacing

{\everypar={\hangindent=0.5truecm}
\parindent=0pt\frenchspacing

~1.~D.W.~Sciama, astro--ph/9811172; and references therein;  A. Geiser, 
CERN--EP/98--56, hep--ph/9901433;  W. Kr\'{o}likowski, hep--ph/9903209 
(Appendix).

\vspace{0.15cm}

~2.~W. Kr\'{o}likowski, hep--ph/9811421; and references therein.

\vspace{0.15cm}

~3.~{\it Cf. e.g.}, J.N. Bahcall, P.I. Krastov and A.Y. Smirnov, hep--ph/%
9807216v2.

\vspace{0.15cm}

~4.~Y. Fukuda {\it et al.} (\UK Collaboration), {\it Phys. Rev. Lett.} {\bf 
81}, 1562 (1998); and references therein. 

\vspace{0.15cm}

~5.~C.~Athanassopoulos {\it et al.} (LSND Collaboration), {\it Phys. Rev.}
{\bf C 54}, 2685 (1996); {\it Phys. Rev. Lett.} {\bf 77}, 3082 (1996); nucl--%
ex/9709006.

\vspace{0.15cm}

~6.~L. Baudis {\it et al.}, hep--ex/9902014. 

\vspace{0.15cm}

~7.~{\it Cf. e.g.}, W. Kr\'{o}likowski, {\it Acta Phys. Pol.} {\bf B 30}, 227 
(1999).

\vfill\eject

\end{document}